\documentstyle[epsf,referee]{mn}

\begin{document}
\def\lsun{{\rm L_{\odot}}}
\def\msun{{\rm M_{\odot}}}
\def\rsun{{\rm R_{\odot}}}
\def\be{\begin{equation}}
\def\ee{\end{equation}}

\title{Analytical solution for the evolution of a binary with stable
mass transfer from a giant}
\author[H. Ritter]{H. Ritter\\ Max--Planck--Institut f\"ur Astrophysik, 
Karl--Schwarzschild--Str. 1, D--85740 Garching, Germany\\}

\newcommand{\lta}{\la}
\newcommand{\gta}{\ga}

\maketitle
\begin{abstract}

We derive a simple analytical solution for the evolution of a close 
binary with nuclear time-scale driven mass transfer from a giant. 
This solution is based on the well-known fact that the luminosity and
the radius of a giant scale to a good approximation as simple power 
laws of the mass $M_{\rm c}$ of the degenerate helium core. Comparison
with results of numerical calculations by Webbink, Rappaport \&
Savonije (1983) show the analytical solution and the power law
approximation to be quite accurate. The analytical solution presented
does also allow (in parametrized form) for non-conservative mass 
transfer. Furthermore it is shown that the near constancy of the mass 
transfer rate over most of the mass transfer phase seen in the results
by Webbink, Rappaport \& Savonije is not a generic feature of this
type of evolution but rather a consequence of a particular choice of
parameters. The analytical solution also demonstrates that the level
of mass transfer is largely set by the core mass of the giant at the
{\em onset} of mass transfer. Finally we show that the model is
selfconsistent and discuss its applicability to low-mass X-ray 
binaries.

\end{abstract}

\section{INTRODUCTION}
\label{sect1}

Stable mass transfer on to a (compact) accretor which is driven by the 
nuclear evolution of a giant is a process of considerable
astrophysical interest. This type of evolution was first encountered 
in numerical simulations by (among others) Kippenhahn, Kohl \& Weigert
(1967) and then studied in some detail by Refsdal \& Weigert (1969, 
1971) in the context of the formation of low-mass white dwarfs in 
binaries. Later Webbink, Rappaport \& Savonije (1983, hereafter WRS)
and, independently, Taam (1983) have proposed that some of the most 
luminous among the galactic low-mass X-ray binaries could be powered
by nuclear time-scale driven mass transfer from a giant. Finally with
the detection of the millisecond pulsar binary PSR 1953+29 by
Boriakoff, Buccheri \& Fauci (1983) it became immediately clear that
this system (and many other similar ones which have been found since
then) must be the end product of an evolution of the type described
by WRS and Taam (1983), see e.g. Joss \& Rappaport (1983), Paczynski
(1983), Savonije (1983), and Rappaport et al. (1995, hereafter RPJDH) 
for a more recent reference.  

Whereas Taam (1983) performed full stellar structure calculations WRS
took a simpler approach by making use of the well-known fact that
giants obey a core mass-luminosity and a core mass-radius relation. 
WRS derived these relations from full stellar equilibrium models of
giants and approximated them as power series of the mass of the
degenerate helium core $M_{\rm c}$. By using the core mass-luminosity
and core mass-radius relation calculating the evolution of a
binary with nuclear time-scale driven mass transfer from a giant
reduces to preforming a simple integral in time which, however, in the
approach taken by WRS has still to be performed numerically. The
purpose of this paper is first to show that the evolution of such a 
binary can be described by an analytical solution if the core 
mass-luminosity and core mass-radius relation are simple power laws 
of $M_{\rm c}$, and second to show that the results of full stellar 
structure calculations of giants given by WRS can indeed 
be approximated with little loss of accuracy by simple power laws. 
Moreover, the analytical solutions which we present are more general 
than the numerical ones of WRS in the sense that we allow for 
non-conservative mass transfer and a non-zero mass radius exponent 
of the donor star (see below section \ref{sect3}). 
 
The advantages of having an analytical solution as compared to a
numerical one are immediatedly clear: First, if sufficiently simple
(as in our case), such solutions provide us with direct physical 
insight into how this type of evolution works, i.e. which parameters 
determine the characteristc properties and which ones are of minor
importance. Second, whereas a numerical solution has
to be computed separately for each set of input parameters of
interest, an analytical solution provides us with the whole manifold
of solutions, i.e. we get the explicit dependence of the solution on 
the input parameters. Among other things, this allows us to examine 
whether certain properties found in numerical solutions are generic or
a simple consequence of a particular choice of parameters. A case in 
point is the well-known property of the solutions presented by WRS, 
namely that the mass transfer rate stays almost constant over most of 
the mass transfer phase. 

Another area of application in which simple analytical solutions are 
superior to numerical ones is population synthesis. This is because 
population synthesis calculations usually involve integrals over the 
complete manifold of solutions of the particular evolutionary phase 
under consideration. Using numerical solutions for that task requires 
computation of a sufficiently dense grid of evolutions over a 
multi-dimensional parameter space and such computations can easily 
become prohibitively expensive (see e.g. Kolb 1993 for a detailed 
discussion in the context of cataclysmic variables). Using analytical 
solutions instead is comparatively easy and cheap. 

This paper is organized as follows: In section \ref{sect2} we present
the basic model assumptions in the framework of which we shall work,
in section \ref{sect3} we give an analytical solution for the nuclear 
evolution of a giant, and in section \ref{sect4} we derive expressions
for the mass radius exponent of the donor star's critical Roche
radius. Finally, in section \ref{sect5} we present the analytical
solution for nuclear time-scale driven mass transfer from a giant. In
order to compare our results with those of WRS we derive in section
\ref{sect6} the parameters of the power law approximations of the core
mass-luminosity and core mass-radius relations using the data from
WRS and RPJDH. In section \ref{sect7} we then compare the results
obtained from the analytical solutions with the corresponding
numerical ones of WRS. In section \ref{sect8} we examine whether the 
model is selfconsistent and whether it is applicable to low-mass X-ray 
binaries. A discussion based on the analytical solution of the key 
features of nuclear time-scale driven mass transfer from a giant
follows in section \ref{sect9}, and our main conclusions are
summarized in the final section \ref{sect10}.

\section{BASIC MODEL ASSUMPTIONS}
\label{sect2}

In the following we shall list the basic model assumptions in the
framework of which we shall work. These assumptions are:

\begin{enumerate}

\item The donor star (of mass $M_2$) is on the first giant branch and
      has a degenerate helium core of mass $M_{\rm c}$. Nuclear luminosity 
      comes from hydrogen shell burning exclusively. 

\item $M_2$ is small enough that the binary is adiabatically and
      thermally stable against mass transfer. For this to hold we must
      have (e.g. Ritter 1988)
\begin{equation}
\zeta_{\rm{ad}} - \zeta_{\rm{R,2}} > 0
\label{2.1}
\end{equation}
      and 
\begin{equation}
\zeta_{\rm{e}} - \zeta_{\rm{R,2}} > 0,
\label{2.2}
\end{equation}
      where 
\begin{equation}
\zeta_{\rm{ad}} = {\biggl(\frac{\partial \ln R_2}{\partial \ln M_2}
\biggr)}_{\rm{ad}}
\label{2.3}
\end{equation} 
      is the adiabatic mass radius exponent of the donor star, 
\begin{equation}
\zeta_{\rm{e}} = {\biggl(\frac{\partial \ln R_2}{\partial \ln M_2}
\biggr)}_{\rm{e}}
\label{2.4}
\end{equation}
      the corresponding thermal equilibrium mass radius exponent, and 
\begin{equation}
\zeta_{\rm{R,2}} = {\biggl(\frac{\partial \ln R_{R,2}}{\partial \ln M_2}
\biggr)}_{\ast}
\label{2.5}
\end{equation}
      the mass radius exponent of its critical Roche radius (see
      e.g. Ritter 1996 for details). In (\ref{2.5}) the subscript
      $\ast$ indicates that for computing this quantity one has to
      specify how, as a consequence of mass transfer,  mass and
      orbital angular momentum are lost from and redistributed within 
      the binary system (see below section \ref{sect4}).

      For stars on the first giant branch and with small relative core
      mass $M_{\rm c}/M$ $\zeta_{\rm{ad}} \approx
      -1/3$. $\zeta_{\rm{ad}}$ increases with relative core mass and
      becomes positive if $M_{\rm{c}}/M  \ga 0.2$ (Hjellming \& Webbink
      1987). Therefore, as we shall see, adiabatic stability of mass 
      transfer is less critical than thermal stability. 

      Assuming that the donor obeys a core mass-luminosity relation,
      i.e. that $L = L(M_{\rm c})$ and $\partial L/\partial M = 0$, 
      where $L$ is the (nuclear) luminosity, $\zeta_{\rm e}$ follows
      from the Stefan--Boltzmann law: 
%
\begin{equation}
\zeta_{\rm e}  =  \frac{1}{2}\, {\left(\frac{\partial \ln L}{\partial 
                    \ln M}\right)}_{\rm e} - 2\, \left(\frac{\partial 
                    \ln T_{\rm eff}}{\partial \ln M}\right)_{\rm e} 
               =  -2\, {\left(\frac{\partial \ln T_{\rm eff}}{\partial 
                    \ln M}\right)}_{\rm e}  
\label{2.6}
\end{equation}
      It is well known that a giant, as long as it has a sufficiently
      deep outer convective envelope, in the HRD stays close to and 
      evolves along the Hayashi line to higher luminosity. This
      property allows us to get an approximation for $\zeta_{\rm e}$. 
      Approximating the Hayashi line in the HRD in the form
      (e.g. Kippenhahn \& Weigert 1990)
\begin{equation}
\log L = a \log T_{\rm eff} + b \log M + c
\label{2.7} 
\end{equation} 
      this together with (\ref{2.6}) yields  
\begin{equation}
\zeta_{\rm e} \approx - 2 b/ a\,.
\label{2.8}
\end{equation}
      Numerical calculations typically yield $-0.3 \la \zeta_{\rm e}
      \la -0.2$. Thus, except for small relative core mass,
      $\zeta_{\rm e} < \zeta_{\rm ad}$, i.e. the criterion for thermal
      stability (\ref{2.2}) against mass transfer is stronger than the 
      one for adiabatic stability (\ref{2.1}). 
\item Mass transfer is driven only by the donor's nuclear evolution,
      i.e. we assume that there is no systemic angular momentum loss 
      in the absence of mass transfer. Consequential angular momentum
      loss (see e.g. King \& Kolb 1995 for a discussion) is, however,
      not excluded a priori. 
\item Mass transfer is so slow that the donor star remains close to
      thermal equilibrium. Accordingly we assume that its radius $R_2$
      and luminosity $L_2$ are given by the corresponding thermal
      equlilibrium values, i.e. $R_2 = R_{2,{\rm e}}$ and $L_2 =
      L_{2,{\rm e}}$, and hence $\zeta_{\rm eff} = (d \ln R_2/d \ln
      M_2) = \zeta_{\rm e}$. This condition is well fulfilled 
      because mass transfer occurs on the nuclear time scale
      $\tau_{\rm nuc}$ which is much longer than the Kelvin-Helmholtz
      time $\tau_{\rm KH} = G {M_2}^2/R_2 L_2$. We shall later (section
      \ref{sect8}) verify the validity of this assumption. 
\end{enumerate}
       
With the above assumptions we can now write the mass transfer rate,
i.e. the donor's mass loss rate as (e.g. Ritter 1996) 
\begin{equation}
-\dot{M}_2 = \frac{M_2}{\zeta_{\rm e} - \zeta_{\rm R,2}}
\left(\frac{\partial \ln R_2}{\partial t}\right)_{\rm nuc},
\label{2.9}
\end{equation}
where $(\partial \ln R_2/\partial t)_{\rm nuc}$ is the inverse
time-scale of the donor's expansion due to nuclear evolution. 
Next we shall work out in more detail $(\partial \ln R_2/
\partial t)_{\rm nuc}$ and $\zeta_{\rm R,2}$.

\section{NUCLEAR EVOLUTION OF A GIANT}
\label{sect3}

Here we are going to exploit the fact that to a very good
approximation the luminosity $L$ of a star on the first giant branch
scales as a simple power law of the mass $M_{\rm c}$ of the degenerate
helium core but does not depend on its total mass (Refsdal \& 
Weigert 1970). Accordingly we make the following ansatz for $L$:
\begin{equation}
L(M_{\rm c}) = L_0 {\left(\frac{M_{\rm c}}{\msun}\right)}^\lambda
\label{3.1}
\end{equation}
It then follows from Eqs. (\ref{2.6}) and (\ref{2.7}) that the radius
of such a star cannot at the same time depend only on the core mass
$M_{\rm c}$ but must also depend on its total mass $M$. Because along
the Hayashi line the effective temperature remains almost constant it
follows from (\ref{3.1}) and the Stefan--Boltzmann law that the
appropriate form of the `core mass-radius relation' is  
\begin{equation}
R(M_{\rm c}, M) = R_0 {\left(\frac{M_{\rm c}}{\msun}\right)}^\rho 
                      {\left(\frac{M}{\msun}\right)}^{\zeta_{\rm e}}.
\label{3.2}
\end{equation}
Here, $L_0$, $\lambda$, $R_0$ and $\rho$ are parameters which have to
be determined from full numerical calculations. We shall give examples
(taken form WRS and RPJDH) in section \ref{sect6}. Using homology 
techniques Refsdal \& Weigert (1970) could derive an expression for 
$\lambda$, the numerical value of which depends on the conditions in 
the hydrogen shell source. A typical value for Pop. I giants is
$\lambda \approx 7 - 8$. 

Hydrogen burning in the shell source around the degenerate core adds
to the core mass at a rate $\dot M_{\rm c}$. Thereby the nuclear
luminosity generated is 
\begin{equation}
L =L_{\rm nuc} = X Q \dot M_{\rm c}\,,
\label{3.3}
\end{equation}
where $X$ is the hydrogen mass fraction in the star's envelope and
$Q \approx 6~10^{18}$ erg g$^{-1}$ the net energy yield of hydrogen
burning per unit mass. In writing (\ref{3.3}) we have assumed that the
star is in thermal equilibrium, i.e. that $L =L_{\rm nuc}$ or, for
practical purposes, that $|L_g| \ll L_{\rm nuc}$, where $L_g$ is the
gravo-thermal luminosity. 

(\ref{3.1}) and (\ref{3.3}) together yield a simple differential
equation for $M_{\rm c}(t)$ with the solution
\begin{equation}
M_{\rm c}(t) =  M_{\rm c}(t=0) \left(1 - \frac{t}{t_\infty}
\right)^{1\over{1 - \lambda}},
\label{3.4}
\end{equation}
where 
\begin{equation} 
t_{\infty} = \frac{X Q \msun}{(\lambda - 1) L_0} {\left[\frac
{M_{\rm c}(t=0)}{\msun}\right]}^{1-\lambda}
\label{3.5}
\end{equation}
is the time over which, formally, the core mass reaches infinity. It
is also the characteristic time for nuclear evolution on the first
giant branch. (\ref{3.4}) combined with respectively (\ref{3.1}) or
(\ref{3.2}) yields 
\begin{equation}
L(t) = L_0\, {\left[ \frac{M_{\rm c}(t=0)}{\msun}\right]}^\lambda 
             {\left(1 - \frac{t}{t_\infty} \right)}^{\frac{\lambda}
             {1 - \lambda}} 
\label{3.6}
\end{equation}
and 
\begin{equation}
R(t) = R_0\, {\left(\frac{M}{\msun}\right)}^{\zeta_{\rm e}}
             {\left[ \frac{M_{\rm c}(t=0)}{\msun}\right]}^\rho 
             {\left(1 - \frac{t}{t_\infty} \right)}^{\frac{\rho}{1 -
             \lambda}}.
\label{3.7}
\end{equation}
Here $M_{\rm c}(t=0)$ is the core mass at the beginning of the
evolution on the first giant branch. This quantity has also to be
taken from full stellar evolution calculations or to be treated as a
free parameter. 

From (\ref{3.7}) we finally obtain 
\begin{equation}
\left(\frac{\partial \ln R}{\partial t}\right)_{\rm nuc}  = 
\frac{\rho L_0}{X Q \msun} {\left(\frac{M_{\rm c}}
{\msun}\right)}^{\lambda -1} = \frac{\rho}{\lambda - 1}~ 
\frac{1}{t_\infty - t}\,,
\label{3.8}
\end{equation}
the expression to be inserted in (\ref{2.9}).

\section{COMPUTING $\zeta_{\rm R,2}$}
\label{sect4}

Here we follow standard arguments which have been given many times in
the literature (e.g. Soberman, Phinney \& van den Heuvel 1997 and 
references therein). Therefore, we shall give here only the
definitions we use and the final results. In the context of this
paper, we assume that mass transfer occurs from the secondary and that
the primary (of mass $M_1$) accretes (on average) a fraction $\eta$ of
the transferred mass, i.e. 
\begin{equation}
dM_1 = - \eta dM_2
\label{4.1}
\end{equation} 
and that the mass $d(M_1 + dM_2) = (1 - \eta) dM_2$ leaves the binary
system carrying away the orbital angular momentum 
\begin{equation}
dJ = \nu J \frac{d(M_1 + M_2)}{M_1 + M_2}\,,
\label{4.2}
\end{equation}
where, for the moment at least, $\eta$ and $\nu$ are free parameters,
\begin{equation}
J = {\left( \frac{G {M_1}^2 {M_2}^2 a}{M_1 + M_2}\right)}^{\frac{1}{2}}
\label{4.3.}
\end{equation}
is the orbital angular momentum, and $a$ the orbital
separation. Writing for the secondary's critical Roche radius 
\begin{equation}
R_{\rm R,2} = a f_2(q), 
\label{4.4}
\end{equation}
where 
\begin{equation}
q = \frac{M_1}{M_2}
\label{4.5}
\end{equation}
is the mass ratio, one derives 
\begin{equation}
\zeta_{\rm R,2} = (1 - \eta) \frac{2 \nu + 1}{1 + q} + \frac{2
\eta}{q} - 2 - {\beta}_2 \left(1 + \frac{\eta}{q} \right), 
\label{4.6}
\end{equation}
with 
\begin{equation}
{\beta}_2 = \frac{d \ln f_2}{d \ln q}\,.
\label{4.7}
\end{equation}
As a further simplification we use for $f_2$ in (\ref{4.4}) and
(\ref{4.7}) the approximation (Paczynski 1971) 
\begin{equation}
{f_2}(q) = {\left(\frac{8}{81}\right)}^{1/3} (1 + q)^{-1/3}, ~~~q \ga
1.25
\label{4.8} 
\end{equation}
which yields 
\begin{equation} 
{\beta}_{2}(q) = - \frac{q}{3 (1 + q)}\,.
\label{4.9}
\end{equation}
Inserting (\ref{4.9}) in (\ref{4.6}) finally yields
\begin{equation}
\zeta_{\rm R,2} = \frac{(1 - \eta)(2 \nu + 1)}{1 + q} 
                + \frac{2 \nu}{q} -2 
                + \frac{\eta + q}{3(1 + q)}\,.
\label{4.10}
\end{equation}
In the following we shall also give the results for two specific 
choices of the parameters $\eta$ and $\nu$:

\subsection{Conservative mass transfer}
\label{sect4.1}

Here $\eta = 1$ and $\nu = 0$. From (\ref{4.6}) and (\ref{4.9}) we 
then have 
%
%
%
\begin{equation}
\zeta_{\rm R,2} = \frac{6 - 5 q}{3 q}\,.
\label{4.12}
\end{equation}
\subsection{Isotropic stellar wind from the primary}
\label{sect4.2}

Here we are considering the case that all the transferred mass is
ejected from the primary in an isotropic stellar wind carrying with it
the specific orbital angular momentum of the primary. Accordingly
$\eta =0$ and $\nu =1/q$. Inserted in (\ref{4.6}) together with
(\ref{4.9}) this yields 
%
%
%
\begin{equation}
\zeta_{\rm R,2} = \frac{6 - 5q^2 - 3q}{3q(1 + q)}\,.
\label{4.14}
\end{equation}
\section{EVOLUTION WITH MASS TRANSFER FROM A GIANT}
\label{sect5}
Now that we have determined $(\partial \ln R/\partial t)_{\rm nuc}$
and $\zeta_{\rm R,2}$ in equation (\ref{2.9}) we can solve that
equation. First we are going to give a rather general solution before 
discussing special cases of interest. 

\subsection{General analytical solution}
\label{sect5.1}
General solution in this context means that we allow for a constant
value $\eta \ge 0$, either a constant value of $\nu \ge 0$ or $\nu =
1/q$, and a constant value of $\zeta_{\rm e}$ subject to the condition
(\ref{2.2}). Inserting (\ref{3.8}) and (\ref{4.6}) with (\ref{4.9})
in (\ref{2.9}) yields an ordinary differential equation for
${M_2}(t)$, or rather, as we shall see, for $t(M_2)$ which is solved
by separating the variables $M_2$ and $t$:
\begin{equation}
\int^{M_2}_{M_{\rm 2,i}}\frac{\zeta_{\rm e}-\zeta_{\rm R,2}}{m_2}\,dm_2
= \int^t_0 {\left(\frac{\partial \ln R}{\partial \tau}\right)}_{\rm
nuc}\,d \tau 
\label{5.1}
\end{equation}
The solution of (\ref{5.1}) can be written as follows:
\begin{equation}
t(M_2)=t_{\infty}\,{\left[1-{\left(\frac{M_1}{M_{\rm 1,i}}\right)}^{-p_1}
{\left(\frac{M_2}{M_{\rm 2,i}}\right)}^{-p_2} {\left(\frac{M_1 +
M_2}{M_{\rm 1,i} + M_{\rm 2,i}}\right)}^{-p_3} e^{-p_4\, \frac{M_{\rm
2,i}-M_2}{M_1}} \right]}\,,
\label{5.2}
\end{equation}
where the exponents $p_1$, $p_2$, $p_3$ and $p_4$ are given in Table 1
for various cases of interest. With (\ref{5.2}) the mass transfer
rate can be written as  
\begin{equation}
-\dot M_2 = \frac{1}{\zeta_{\rm e}-\zeta_{\rm R,2}}\, \frac{\rho
L_0}{X Q}\, {\left(\frac{M_{\rm c,i}}{\msun}\right)}^{\lambda -1}\,
\left(\frac{M_2}{\msun}\right)\, {\left(1-\frac{t}{t_{\infty}}
\right)}^{-1}\,.
\label{5.3}
\end{equation}
%
%
%
Inserting (\ref{5.2}) in (\ref{3.6}) and (\ref{3.7}) finally also 
yields
\begin{equation}
L_2 = L_0\,{\left(\frac{M_{\rm c,i}}{\msun}\right)}^{\lambda}\,
\left[{\left(\frac{M_1}{M_{\rm 1,i}}\right)}^{p_1}\,
{\left(\frac{M_2}{M_{\rm 2,i}}\right)}^{p_2}\,
{\left(\frac{M_1 + M_2}{M_{\rm 1,i} + M_{\rm 2,i}}\right)}^{p_3}\,
e^{p_4\,\frac{M_{\rm 2,i} - M_2}{M_1}}\right]^
{\frac{\lambda}{\lambda -1}}
\label{5.5}
\end{equation}
and 
\begin{equation}
R_2 = R_0\,{\left(\frac{M_{\rm c,i}}{\msun}\right)}^{\rho}\,
\left(\frac{M_2}{\msun}\right)^{\zeta_{\rm e}}\,
{\left[{\left(\frac{M_1}{M_{\rm 1,i}}\right)}^{p_1}\,
{\left(\frac{M_2}{M_{\rm 2,i}}\right)}^{p_2}\,
{\left(\frac{M_1 + M_2}{M_{\rm 1,i} + M_{\rm 2,i}}\right)}^{p_3}\,
e^{p_4\,\frac{M_{\rm 2,i} - M_2}{M_1}}\right]}^
{\frac{\rho}{\lambda -1}}\,.
\label{5.6}
\end{equation}
Here $M_{\rm 1,i}$, $M_{\rm 2,i}$ and $M_{\rm c,i}$ are respectively 
the initial values, i.e. taken at $t = 0$, of $M_1$, $M_2$, and 
$M_{\rm c}$. Note that from (\ref{4.1}) we have  
\begin{equation}
M_1 = M_{\rm 1,i} + \eta \left(M_{\rm 2,i} - M_2 \right)\,.
\label{5.7}
\end{equation}
As can be seen the equations (\ref{5.3}) (together with 
(\ref{5.2})), (\ref{5.5}) and (\ref{5.6}) do not depend explicitly on
time $t$. The solutions are completely fixed by the parameters of the 
core mass-luminosity relation (\ref{3.1}), i.e. $L_0$ and $\lambda$,
the `core mass-radius relation' (\ref{3.2}), i.e. $R_0$, $\rho$ and 
$\zeta_{\rm e}$, the initial values and the current values of both 
masses and the secondary's core mass. Kowing $\dot M_2$, $R_2$ and
$L_2$ as explicit functions of the initial values and current values
of $M_1$, $M_2$ and $M_{\rm c}$ other quantities of interest such as 
the orbital period $P$, the orbital separation $a$, the size of the
accretor's Roche radius $R_{\rm R,1}$ and with it the size of the disc
around the accretor and the secondary's effective temperature 
$T_{\rm eff,2}$ follow directly. What this solution does not provide 
explicitly is the total duration of the mass transfer phase., i.e. the 
time $t_{\rm f}$ when the secondary's envelope is exhausted or, which 
is equivalent, the final core mass $M_{\rm c,f}$. A very good estimate 
of the latter together with $t_{\rm f}$ can be obtained by inserting
(\ref{5.2}) with $M_2 = M_{\rm 2,f} \approx M_{\rm c,f}$ in
(\ref{3.4}). Note that the resulting equation for $M_{\rm c,f}$ cannot
in general be solved in closed form. Reinserting the solution of this
equation in (\ref{5.2}) finally yields also $t_{\rm f}$.   
 
\subsection{Special cases}
\label{sect5.2}

Table 1 lists the powers $p_1$, $p_2$, $p_3$ and $p_4$ which appear in
equations (\ref{5.2}) -- (\ref{5.6}) for a number of special cases of 
interest. 

Case 1 is the most general case for constant values of $\eta$, $\nu$
and $\zeta_{\rm e}$. In case 2 the special choice $\nu = 1/q$ with 
$\eta > 0$ is made, whereas case 3 corresponds to case 2 in the
limit $\eta = 0$. Cases 4 and 5 correspond to conservative mass transfer
with respectively $\zeta_{\rm e} = {\rm const.} \ne 0$ and 
$\zeta_{\rm e} = 0$. Case 6 finally corresponds to case 3 with 
$\zeta_{\rm e} = 0$. 

Case 3 (or 6) corresponds to the situation in which the accretor
ejects all the transferred mass with the specific orbital angular
momentum of the accretor's orbit. This approximates what apparently
must happen in systems in which the mass transfer rate to a neutron 
star exceeds the Eddington accretion rate by many orders of magnitude
(see e.g. King \& Ritter 1999 for a discussion). 

We shall later mostly make use of case 5 because this is the
approximation within which WRS have performed their numerical
calculations and with which we wish to compare the results of our
analytical solution. Before we can do that we have to provide the
parameters describing the core mass-luminosity and core mass-radius
relation. 


\begin{table*}
\begin{minipage}{\hsize}
\caption{Parameters and coefficients entering the analytical solution}

\begin{tabular}{rccccccc}\hline
case &      $\eta$   & $\nu$ &$\zeta_{\rm e}$&            $p_1$
&                     $p_2$                     &                $p_3$
&           $p_4$             \\ \hline
 & & & & & & & \\
   1 &$\ge 0$, const.& const.&    const.     &$-\frac{2(\lambda -1)}{\rho}$     &$-\frac{(3\zeta_{\rm e}+5)(\lambda -1)}{3\rho}$&$\frac{2(3\nu +1)(\lambda -1)}{3\rho}$&             0               \\ 
 & & & & & & & \\
   2 &$  > 0$, const.&$1/q$  &    const.     &$-\frac{2(\lambda -1)}{\eta \rho}$&$-\frac{(3\zeta_{\rm e}+5)(\lambda -1)}{3\rho}$&$-\frac{4(\lambda -1)}{3\rho}$        &             0               \\
 & & & & & & & \\
   3 &      0        &$1/q$  &    const.     &              0                   &$-\frac{(3\zeta_{\rm e}+5)(\lambda -1)}{3\rho}$&$-\frac{4(\lambda -1)}{3\rho}$        &$-\frac{2(\lambda -1)}{\rho}$\\
 & & & & & & & \\
   4 &      1        &  ---- &    const.     &$-\frac{2(\lambda -1)}{\rho}$     &$-\frac{(3\zeta_{\rm e}+5)(\lambda -1)}{3\rho}$&                0                     &             0               \\
 & & & & & & & \\
   5 &      1        &  ---- &      0        &$-\frac{2(\lambda -1)}{\rho}$     &$-\frac{5(\lambda -1)}{3\rho}$                 &                0                     &             0               \\
 & & & & & & & \\
   6 &      0        &$1/q$  &      0        &              0
&$-\frac{5(\lambda -1)}{3\rho}$                 &$-\frac{4(\lambda
-1)}{3\rho}$        &$-\frac{2(\lambda -1)}{\rho}$\\ \hline 
\end{tabular}
\end{minipage}
\end{table*}

\section{Approximations for the core mass-luminosity and core mass
radius relation}
\label{sect6}

In general the values of $L_0$, $\lambda$, $R_0$, $\rho$ and 
$\zeta_{\rm e}$ in equations (\ref{3.1}) and (\ref{3.2}) have to be 
determined from full numerical calculations of stars in thermal  
equilibrium on the first giant branch. Here we restrict ourselves to 
deriving approximate values of these parameters by using results given
by WRS (their equations 4 -- 6). Because WRS have assumed  
$\zeta_{\rm e} = 0$ their relation (5) is a true core mass-radius 
relation in the sense that the radius of the star does not depend on 
the total mass $M$. Their results for Pop. I and Pop. II giants are 
shown graphically respectively in Fig. 1 and Fig. 2. It is immediately
apparent that in the range $0.15 \msun \la M_{\rm c} \la 0.4 \msun$
the relations $L(M_{\rm c})$ and $R(M_{\rm c})$ are very
nearly simple power laws in $M_{\rm c}$. Based on the results of
Refsdal \& Weigert (1970) this is of course not surprising. Because
our analytical solutions of section \ref{sect5} are based on the power
law approximations (\ref{3.1}) and (\ref{3.2}) we fit the results of 
WRS shown in Figs. 1 and 2 accordingly. The resulting parameters, 
together with the chemical composition of the underlying numerical 
models (characterized in the usual way by the hydrogen mass fraction 
$X$ and the the metallicity $Z$) are summarized in Table 2. The 
corresponding relations are shown as thin full ($L$) or dashed ($R$) 
lines in Figs. 1 and 2. 

\begin{figure}
\begin{minipage}[t]{0.48\hsize}
 \centerline{\epsfxsize=1.0\hsize\epsffile{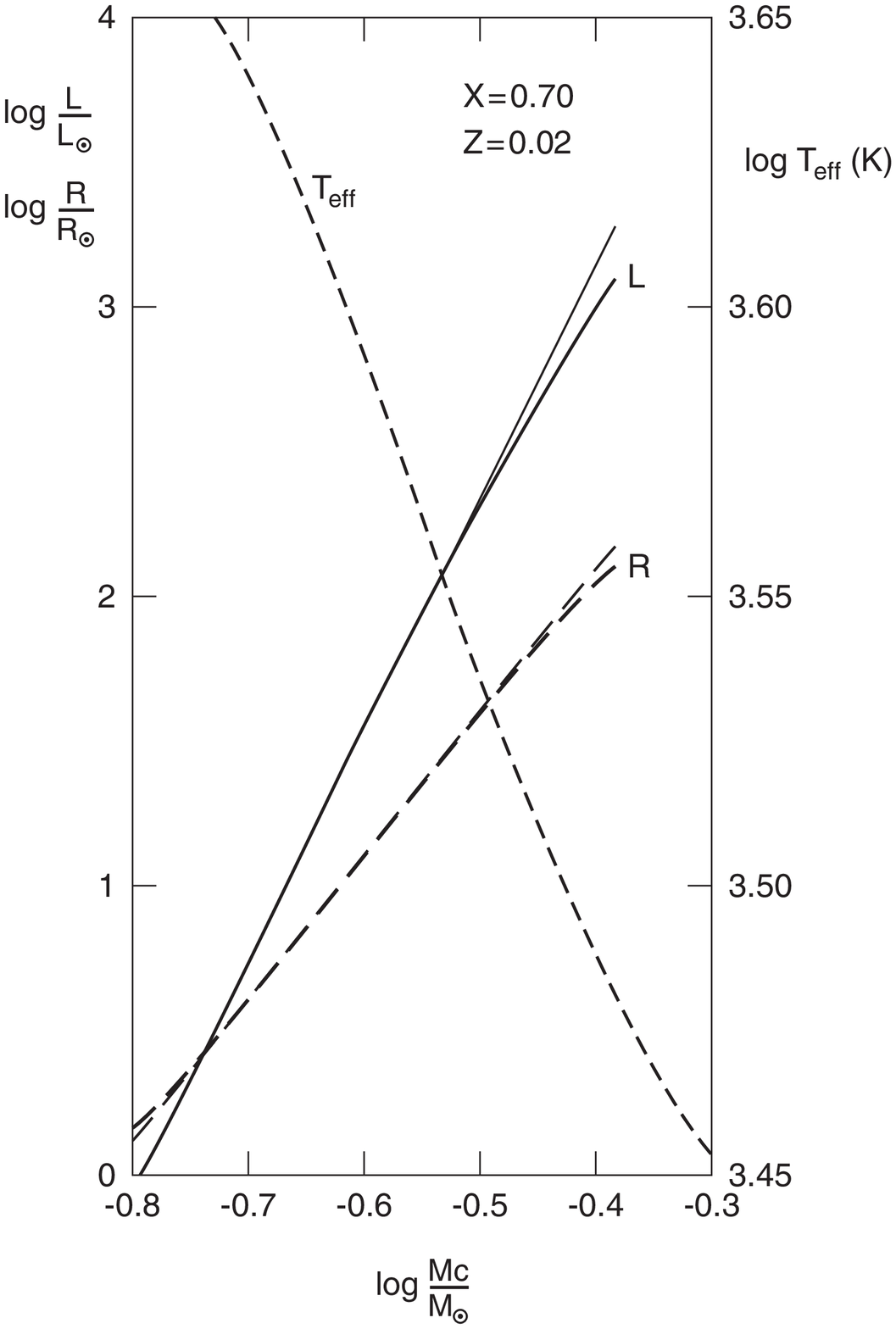}}
\caption{Core mass-luminosity relation (full line), core mass-radius 
relation (long-dashed line) and core mass-effective temperature 
relation (short-dashed line) for Pop. I giants according to numerical
calculations by WRS. Our simple power law fits are shown respectively 
as thin full and thin long-dashed lines. The corresponding fit 
parameters are listed in Table 2.}
\end{minipage}
\hfill
\begin{minipage}[t]{0.48\hsize}
 \centerline{\epsfxsize=1.0\hsize\epsffile{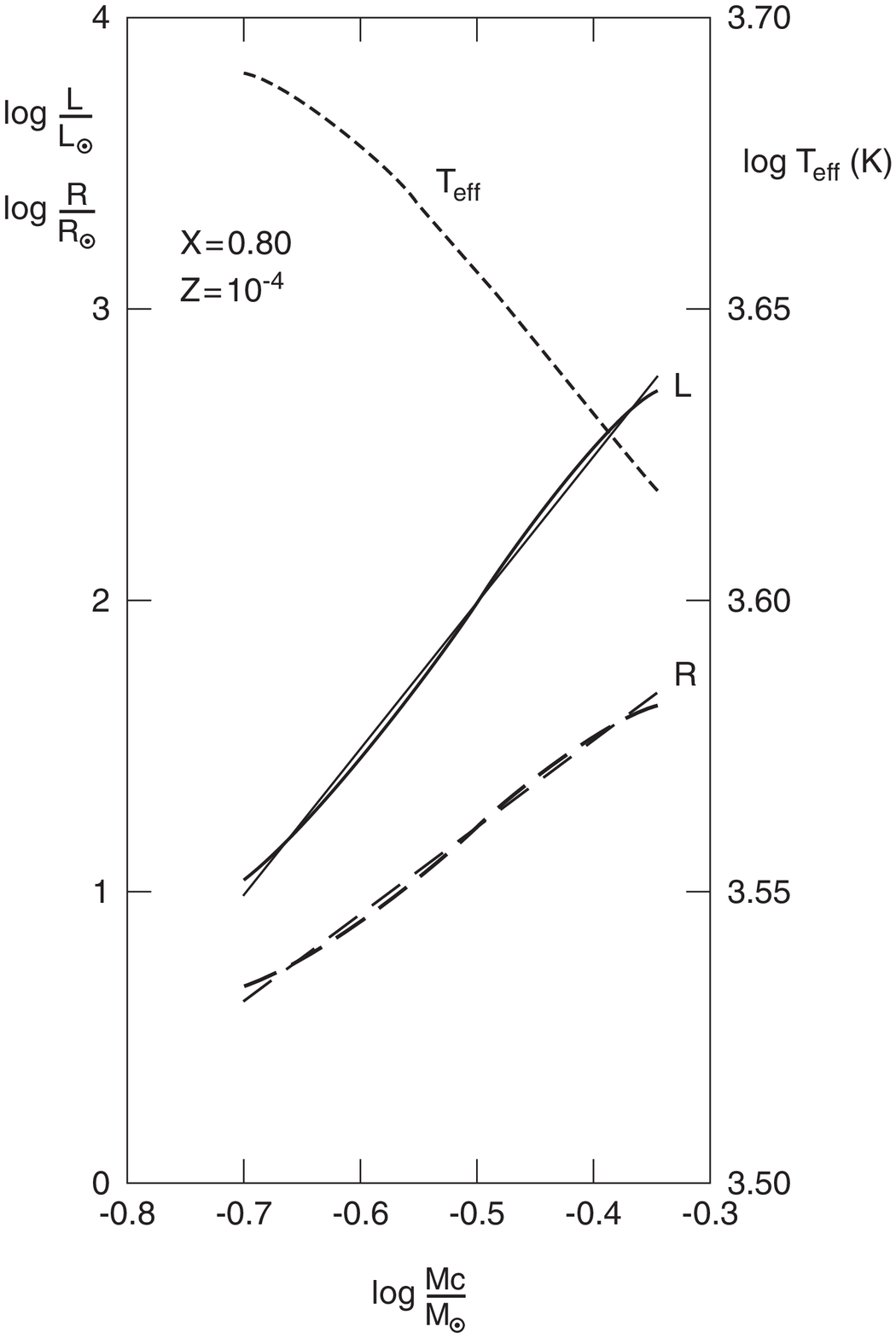}}
\caption{The same as Fig. 1, but for Pop. II giants.} 
\end{minipage}
\end{figure}

For comparison we provide in Table 2 also approximate parameters of a
power law fit in the range $0.2 \msun \la M_{\rm c} \la 0.5 \msun$ to 
the core mass-radius relation derived by RPJDH (their equation 5).  


\begin{table*}
\begin{minipage}{\hsize}
\caption{Approximate parameters of the core mass-luminosity and core
mass-radius relations derived from numerical results}

\begin{tabular}{ccccccccc}\hline
Pop. &  X   &  Z    & $\log\frac{L_0}{\lsun}$ & $\lambda$ & $\log\frac{R_0}{\rsun}$ & $\rho $ & $\zeta_{\rm e}$ & source \\ \hline
     &      &       &                         &           &                         &         &                 &        \\
  I  & 0.70 & 0.02  &          6.3            &     8     &           4.1           &    5    &        0        &  WRS   \\
     &      &       &                         &           &                         &         &                 &        \\
 II  & 0.80 &0.0001 &          4.5            &     5     &           2.7           &    3    &        0        &  WRS   \\
     &      &       &                         &           &                         &         &                 &        \\
  I  & 0.70 & 0.02  &          ---            &    ---    &           3.58          &   4.21  &        0        & RPJDH  \\
     &      &       &                         &           &                         &         &                 &        \\
 II  & 0.70 & 0.001 &          ---            &    ---    &           3.36          &   4.17  &        0        & RPJDH  \\ \hline

\end{tabular}
\end{minipage}
\end{table*}

\section{NUMERICAL RESULTS VERSUS THE ANALYTICAL MODEL}
\label{sect7}

Here we wish to compare the results obtained from the analytical
approximation, i.e. from equations (\ref{5.2}) -- (\ref{5.6}), with
the numerical results obtained by WRS. For this we have to insert 
the parameters listed in Table 2 into the expressions for $p_1$,
$p_2$, $p_3$, and $p_4$ corresponding to case 5 in Table 1 and in
equations (\ref{5.2}) -- (\ref{5.6}). Rather than listing all the
resulting equations for both Pop. I and Pop. II composition, we
restrict ourselves to giving only the corresponding expressions for 
the mass transfer rate. With $Q = 6~10^{18}$ erg g$^{-1}$ this yields 
\begin{equation}
-\dot M_2 = 2.64~10^{-8}\,\msun {\rm yr^{-1}}\,
\left(\frac{M_{\rm c,i}}{0.25\msun}\right)^7\,
\frac{M_1}{5M_1 - 6M_2}\,
\left(\frac{M_2}{\msun}\right)\,
\left(\frac{M_2}{M_{\rm 2,i}}\right)^{-7/3}\,
\left(\frac{M_1}{M_{\rm 1,i}}\right)^{-14/5}
\label{7.1}
\end{equation}
for Pop. I, and 
\begin{equation}
-\dot M_2 = 1.40~10^{-8}\,\msun {\rm yr^{-1}}\,
\left(\frac{M_{\rm c,i}}{0.25\msun}\right)^4\,
\frac{M_1}{5M_1 - 6M_2}\,
\left(\frac{M_2}{\msun}\right)\,
\left(\frac{M_2}{M_{\rm 2,i}}\right)^{-20/9}\,
\left(\frac{M_1}{M_{\rm 1,i}}\right)^{-8/3}
\label{7.2}
\end{equation} 
for Pop. II chemical composition. 

Results of detailed numerical computations by WRS are shown in their
figs. 3 and 4. Comparing the results obtained from our analytical 
solution for the same parameters with those of WRS shows that the 
differences are small, too small in fact to be adequately shown on the
scale of these figures. That the differences in $L(t)$ and $R(t)$ are
small was to be expected based on the quality of the fits used (see
Figs. 1 and 2). That the differences in $-\dot M_2$ are small too is shown
in Figs. 3 and 4. In that context we should emphasize that our fits of
the $L(M_{\rm c})$ and $R(M_{\rm c})$ relations of WRS with power laws
are the simplest possible ones (with integer powers $\lambda$ and
$\rho$) and that no attempt has been made to adjust the power law
parameters such as to minimize the differences between the numerical
results of WRS and the analytical solution. With $-\dot M_2$ being in
close agreement with the numerical results it follows that also
$M_2(t)$ and $P(t)$ should agree very well, which is indeed the case. 
Finally also $M_{\rm c}(t)$ must be in close agreement with the
numerical result of WRS, otherwise the temporal evolution of the other
quantities would not match either. 

\begin{figure}
  \begin{minipage}[t]{0.48\hsize}
  \centerline{\epsfxsize=1.0\hsize\epsffile{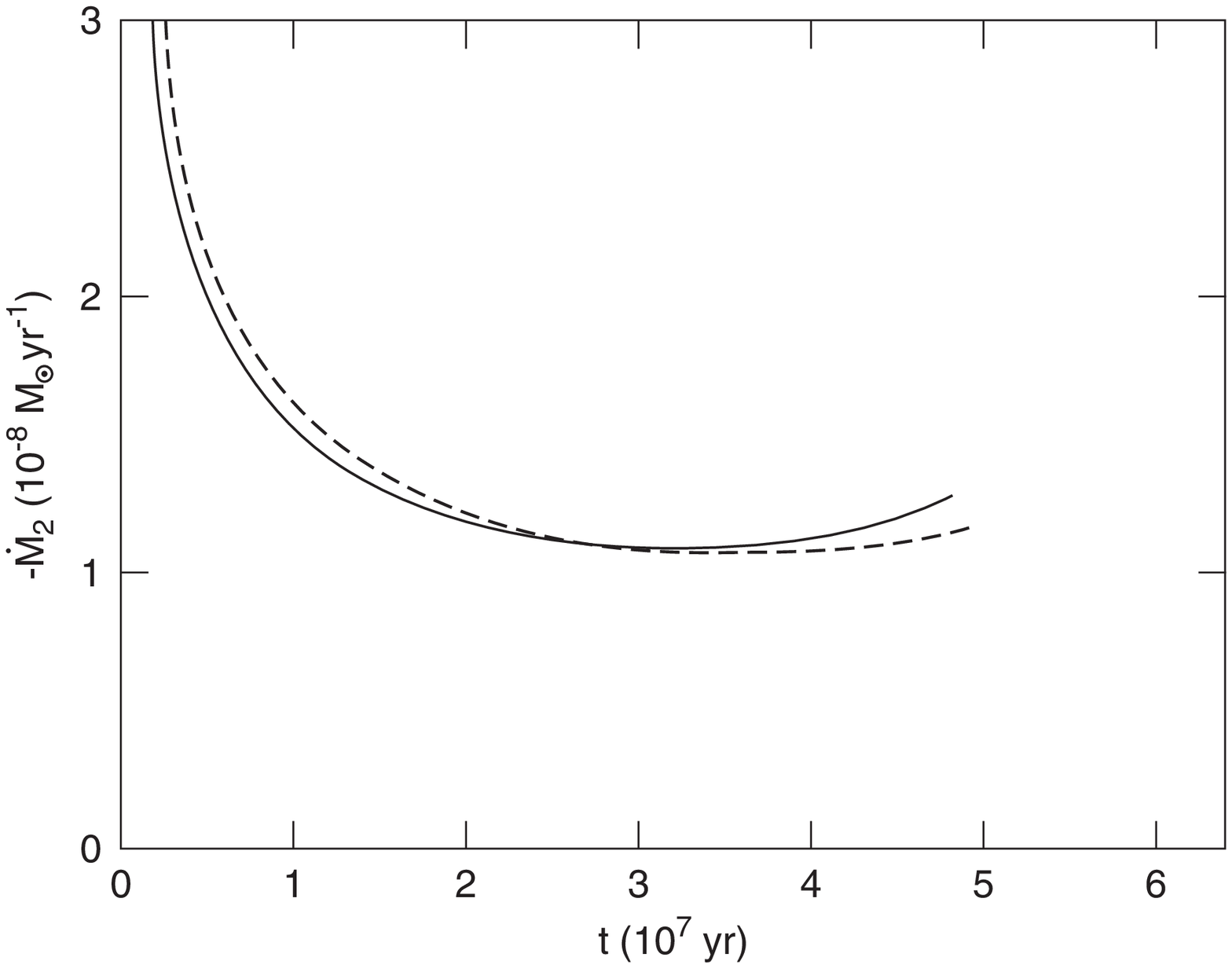}}
\caption{Comparison of the result obtained with the analytical solution (full
line) with that of a matching numerical one by WRS (dashed
line). Shown is the mass loss rate from a Pop. I giant as a function
of the time for the parameters used by WRS in their figure 3, i.e. 
$X = 0.70$, $Z = 0.02$, $M_{\rm 1,i} = 1.4 \msun$, $M_{\rm 2,i} = 1 
\msun$, and $M_{\rm c,i} = 0.26 \msun$.}
  \end{minipage}
  \hfill
  \begin{minipage}[t]{0.48\hsize}
  \centerline{\epsfxsize=1.0\hsize\epsffile{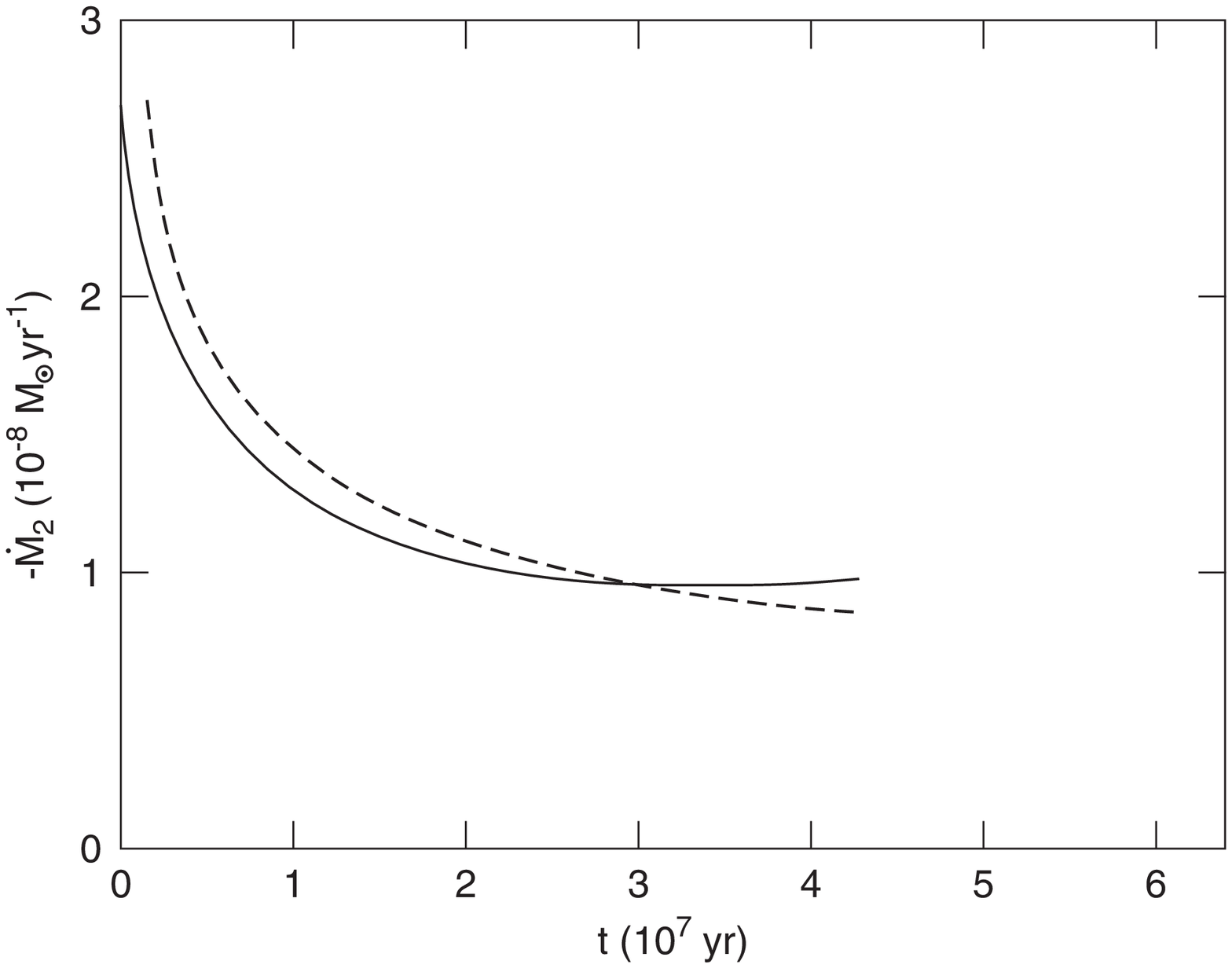}}
\caption{Comparison of the result obtained with the analytical solution (full
line) with that of a matching numerical one by WRS (dashed
line). Shown is the mass loss rate from a Pop. II giant as a function
of the time for the parameters used by WRS in their figure 4, i.e. 
$X = 0.80$, $Z = 10^{-4}$, $M_{\rm 1,i} = 1.4 \msun$, $M_{\rm 2,i} =
0.9 \msun$, and $M_{\rm c,i} = 0.31 \msun$.}
  \end{minipage}
\end{figure}

In fact we can summarize the result of this comparison as follows: if
the power law fits to the core mass-luminosity and core mass-radius 
relations obtained from full stellar structure calculations are 
adequate, the analytical model provides an adequate approximation to the
results of numerical computations. 

\section{SELFCONSISTENCY AND APPLICABILITY OF THE MODEL}
\label{sect8}

\subsection{Deviation from thermal equilibrium}
\label{sect8.1}

The equations we have derived in section \ref{sect5} are valid only to
the extent that the donor star is not significantly driven out of
thermal equilibrium as a consequence of mass loss. That is to say that
the time-scale of mass loss 
\begin{equation}
\tau_{\rm M_2} = -\frac{dt}{d \ln M_2}
\label{8.1}
\end{equation}
must be significantly longer than the shortest time-scale on which 
the donor's convective envelope can react, i.e. longer than the 
Kelvin--Helmholtz time of the convective envelope (e.g. King et al. 
1997a)  
\begin{equation}
\tau_{\rm KH,ce} = \frac{3}{7}\, \frac{M_{\rm ce}}{M}\, \tau_{\rm KH} 
                 = \frac{3}{7}\, \frac{M - M_{\rm c}}{M}\, 
                   \frac{G M^2}{R L}\,.
\label{8.2}
\end{equation}
Here $M_{\rm ce} = M - M_{\rm c}$ is the mass of the convective
envelope and $\tau_{\rm KH}$ the Kelvin--Helmholtz time of the whole
star. A measure for the star's deviation from thermal equilibrium is 
thus the the ratio $\tau_{\rm M_2}/\tau_{\rm KH,ce}$. Using equations 
(\ref{2.9}) and (\ref{3.6}) -- (\ref{3.8}) in (\ref{8.1}) and 
(\ref{8.2}) we can now write 
\begin{equation}
\tau_{\rm M_2} = \left(\zeta_{\rm e} - \zeta_{\rm R,2}\right)\,
                 \frac{X Q \msun}{\rho L_0}\,
                 \left(\frac{M_{\rm c,i}}{\msun}\right)^{1-\lambda}\,
                 \left(1 - \frac{t}{t_{\infty}}\right)
\label{8.4}
\end{equation}
and
\begin{equation}
\tau_{\rm KH,ce} = \frac{3}{7}\, \frac{G {\msun}^2}{R_0 L_0}\,
                   \frac{M_2 - M_{\rm c}}{M_2}\,
                   \left(\frac{M_2}{\msun}\right)^{2-\zeta_{\rm e}}\,
                   \left(\frac{M_{\rm c,i}}{\msun}\right)^{-\left(
                   \lambda + \rho \right)}\,
                   \left(1 - \frac{t}{t_{\infty}}\right)^{\frac{\lambda +
                   \rho}{\lambda -1}}\,,
\label{8.5}
\end{equation}
and, therefore, 
\begin{equation}
\frac{\tau_{\rm M_2}}{\tau_{\rm KH,ce}} =
   \left(\zeta_{\rm e}-\zeta_{\rm R,2}\right)\,
   \frac{7}{3}\, \frac{X Q R_0}{\rho G \msun}\,
   \frac{M_2}{M_2 - M_{\rm c}}\,
   \left(\frac{M_2}{\msun}\right)^{\zeta_{\rm e}-2}\,
   \left(\frac{M_{\rm c,i}}{\msun}\right)^{1+\rho}\,
   \left(1 -\frac{t}{t_{\infty}}\right)^{\frac{1+\rho}{1-\lambda}}\,.
\label{8.6}
\end{equation}
Inserting now for illustrative puposes the parameters of Table 2 which
correspond to the results of WRS into (\ref{8.6}) yields 
\begin{equation}
\frac{\tau_{\rm M_2}}{\tau_{\rm KH,ce}} = 3.16~10^3\,
   \left(-\zeta_{\rm R,2}\right)\, \frac{M_2}{M_2 - M_{\rm c}}\,
   \left(\frac{M_2}{\msun}\right)^{-2}\,
   \left(\frac{M_{\rm c,i}}{0.25~\msun}\right)^6\,
   \left(1 - \frac{t}{t_{\infty}}\right)^{-6/7}
\label{8.7}
\end{equation}
for Pop. I, and 
\begin{equation}
\frac{\tau_{\rm M_2}}{\tau_{\rm KH,ce}} = 3.83~10^3\,
   \left(-\zeta_{\rm R,2}\right)\, \frac{M_2}{M_2 - M_{\rm c}}\,
   \left(\frac{M_2}{\msun}\right)^{-2}\,
   \left(\frac{M_{\rm c,i}}{0.25~\msun}\right)^4\,
   \left(1 - \frac{t}{t_{\infty}}\right)^{-1}
\label{8.8}
\end{equation}
for Pop. II chemical composition. 

What (\ref{8.6}) -- (\ref{8.8}) demonstrate is that for all practical
purposes $\tau_{\rm M_2}/\tau_{\rm KH,ce} \gg 1$ unless the mass ratio
is such that the binary system is close to thermal instability,
i.e. that $\zeta_{\rm e} - \zeta_{\rm R,2} \ll 1$. Thus  mass transfer
driven by nuclear evolution of a star on the first giant branch is
always sufficiently slow for the donor star to remain essentially in
thermal equilibrium. To this extent, the analytical model presented in
section \ref{sect5} is selfconsistent. 

\subsection{Irradiation of the donor star}
\label{sect8.2}

Since one of the main areas of application of this model is the
evolution of luminous X-ray binaries (see e.g. Taam 1983 and WRS) and
their transformation into millisecond pulsar binaries (see
e.g. RPJDH and references therein) we have to examine
whether this model remains applicable if irradiation of the cool donor
star by X-rays from the vicinity of a very compact accretor is taken
into account. As has first been shown by Podsiadlowski (1991) cool
stars, when exposed to sufficiently strong external heating by
X-rays, respond by expansion on the thermal time-scale of the 
convective envelope, thereby giving rise to very high mass transfer 
rates in a semi-detached binary. In the context of mass transfer from
a giant to a compact accretor the stability of mass transfer with 
taking into account irradiation has been examined in some detail by 
King et al. (1997a). Their finding was that if the accretion rate on to
the compact star equals the mass transfer rate at any time, the
evolutionary model discussed here is not in general applicable,
because mass transfer is very likely to be unstable. Rather than being 
continuous, mass transfer occurs in cycles which are characterized by 
short high states during which very high mass transfer rates are  
achieved and long low states during which the binary is detached. 
As has been shown by King et al. (1997a) this instability occurs
mainly because for giants the ratio of mass loss time-scale to
Kelvin--Helmholtz time (equation \ref{8.6}) is so large. It is
therefore somewhat ironic that self-consistency of the model of
nuclear time-scale driven stationary mass transfer from a giant is
possible only at the price that such systems are in danger of being 
unstable to irradiation-driven mass transfer. 

However, the above-mentioned condition, namely that the accretion
rate equals the mass transfer rate at any time, is not likely to be 
fulfilled. Rather, as has been shown by King et al. (1997b), the
accretion disc in such binary systems is likely to be unstable and
that even for the rather optimistic assumptions made by King et
al. (1997b) about the efficiency of X-ray irradiation of the
accretion disc. Based on newer results of Dubus et al. (1998) one
could conclude that the accretion discs in essentially all systems
with a giant donor are unstable. If this is the case, then accretion
on to the compact star is spasmodic because the disc undergoes a dwarf
nova-like limit-cycle instability. Estimates of the properties of
the outbursts (e.g. King \&  Ritter 1998) show that they are
characterized by a very small duty cycle. This, in turn, is now of
great importance for the stability of mass transfer from the giant in
the presence of X-ray irradiation: King et al. (1997a) show that if
accretion on to a neutron star or a black hole is spasmodic with a
small duty cycle, then mass transfer from the giant is stable despite 
the donor being irradiated. If that is the case, the model considered 
by Taam (1993), WRS and us is applicable. 

On the other hand, if the accretor is a white dwarf the situation is
more complicated: If the core mass of the donor is $M_{\rm c} \la 0.2
\msun$, irradiation cannot destabilize mass transfer and, therefore, 
the model is applicable. Yet if $M_{\rm c} \ga 0.2 \msun$, mass
transfer is unstable. Although spasmodic accretion has a stabilizing 
effect, for mass transfer to become stable in this case the required 
small duty cycle has to be smaller still by a factor of $\sim 10 
\cdots 100$ than in the case of a black hole or a neutron star (see 
King et al. 1997a, fig. 4). Because the value of the duty cycle does 
not depend explicitely on the nature of the accretor but rather on the
average mass transfer rate and on the size of the accretion disc, it 
appears unlikely that the duty cycle in a white dwarf system is 
systematically smaller than in a black hole or neutron star system. 
Therefore it is more likely that in a white dwarf system in which 
$M_{\rm c} \ga 0.2 \msun$ mass transfer is unstable despite spasmodic 
accretion. In this case the model in question is not applicable. 
However, we should emphasize that detailed numerical calculations of 
mass transfer from a giant with taking into account the effects of 
irradiation of the donor star and of disc instabilites have yet to be 
done. Before results of such computations are available, a final 
decision about whether this model is applicable cannot be made.

\section{DISCUSSION}
\label{sect9}

Having established that the analytical approximation is both accurate  
and selfconsistent, and that the model is probably also applicable to 
the most important case of interest, i.e. to low-mass X-ray
binaries, we can now proceed to discuss the properties of 
these solutions. One of the remarkable properties of this type of
evolution which has already been pointed out by WRS is that the mass 
transfer rate remains almost constant for most of the mass transfer 
phase. Here we shall examine whether this is a generic property of
mass transfer from a giant or whether this is due to a particular 
choice of parameters. Other aspects which we wish examine are the 
dependence of the temporal evolution on the initial parameters, in 
particular on the initial core mass $M_{\rm c,i}$, and systematic
differences between Pop. I and Pop. II systems. 

\begin{figure}
  \centerline{\epsfxsize=0.5\hsize\epsffile{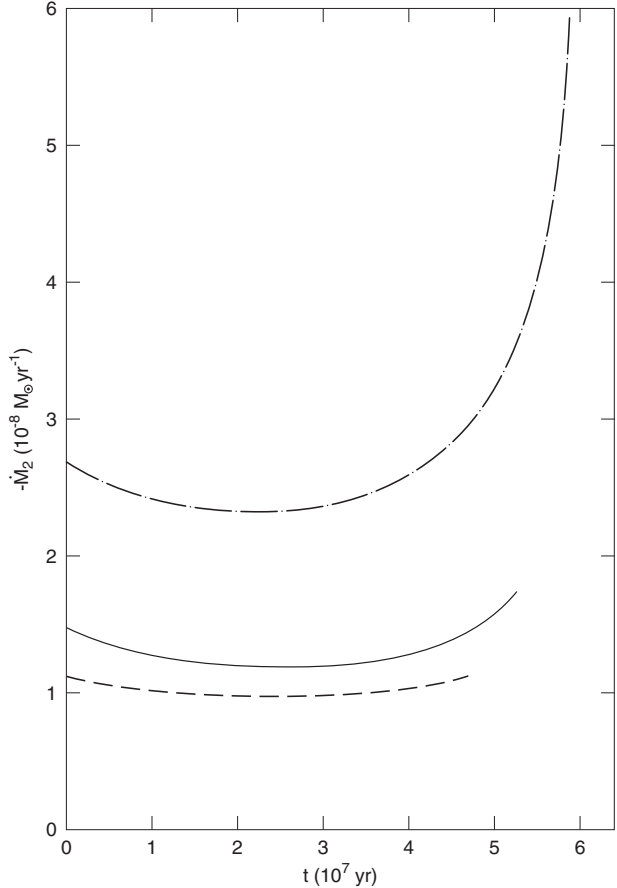}}
\caption{Mass transfer rate computed from the analytical solution for three
different cases: The full and dashed lines show respectively the 
mass transfer rate as a function of time in an evolution of a Pop. I 
and a Pop. II system with same initial parameters as those used in 
Figs. 3 and 4, but now for non-conservative mass transfer 
characterized by $\eta = 0$ and $\nu = 1/q$. The dash-dotted line 
shows the mass transfer rate in a conservative evolution of a Pop. I 
system with the parameters $M_{\rm 1,i} = 5 \msun$, $M_{\rm 2,i} = 2 
\msun$, and $M_{\rm c,i} = 0.26 \msun$.}
\end{figure}

\subsection{The mass transfer rate as a function of time}
\label{sect9.1}

One of the remarkable properties of the evolutions shown by WRS in
their figs. 3 and 4 (see also our Figs. 3 and 4) is that the mass
transfer rate remains constant within a small factor ($\lta 2$) in
the case of conservative mass transfer, and, as Fig. 5 shows, within
an even smaller factor ($\lta 1.5$) in the case that $\eta = 0$ and
$\nu = 1/q$ throughout the evolution. This raises the question why
this is so and whether this is a generic property of this type of
evolution. For our discussion we combine (\ref{2.9}) and (\ref{3.8})
to write the mass transfer rate as 
\begin{equation}
-\dot M_2 = \frac{M_2}{\zeta_{\rm e} - \zeta_{\rm R,2}}\,
            \frac{\rho}{\lambda -1}\, 
            \frac{1}{t_{\infty} - t}\,.
\label{9.1}
\end{equation}
From (\ref{9.1}) it is immediately seen, that it contains two
counteracting factors: first, with time $\zeta_{\rm e} - \zeta_{\rm
R,2}$ increases because q increases. This factor has thus the tendency
to reduce $-\dot M_2$ with time. On the other hand, $t_{\infty} - t$
decreases with time and thus has the tendency to increase $-\dot
M_2$. Therefore, there is a possibility that $-\dot M_2$ goes through 
a minimum during an evolution. Yet, an evolution does not necessarily 
include this minimum. Rather, whether or not
$-\dot M_2(t)$ has a minimum depends on the initial mass ratio $q_i$
and on the subsequent change of $q$ with time. Starting from 
equations (\ref{5.2}) and (\ref{5.3}) and working out $\ddot M_2 = 0$
leads to a (rather complicated) equation for the mass ratio $q_0$ at
which $-\dot M_2$ becomes extremal. From the above discussion we know
already that this extremum is a minimum. Denoting the final mass 
ratio by $q_{\rm f} = M_{\rm 1,f}/M_{\rm 2,f} \approx M_{\rm 1,f}/
M_{\rm c,f}$, we can now distinguish three different cases: 1) if 
$q_{\rm crit} < q_{\rm i} < q_0 < q_{\rm f}$ then the mass transfer 
rate goes through a minimum, 2) if $q_{\rm crit} < q_0 < q_{\rm i} <
q_{\rm f}$, the mass transfer rate will increase monotonically, and if
3) $q_{\rm crit} < q_{\rm i} < q_{\rm f} < q_0$, the mass transfer
rate will decrease monotonically. Here $q_{\rm crit}$ is the critical
mass ratio for both thermal and adiabatic stability against mass
transfer. Its value is given by the larger of the positive solutions 
of the equations $\zeta_{\rm ad} - \zeta_{\rm R,2} = 0$ and 
$\zeta_{\rm e} - \zeta_{\rm R,2} = 0$ (see e.g. Ritter 1988). 

For illustrative purposes we give the values of $q_{\rm crit}$ and
$q_0$ for the cases we show in  Figs. 3 -- 5. For these cases we have
assumed with WRS $\zeta_{\rm e} = 0$ and implicitly $\zeta_{\rm e} < 
\zeta_{\rm ad}$. From this and (\ref{4.12}) it follows that $q_{\rm
crit} = 1.2$ in the case of conservative mass transfer, and with
(\ref{4.14}) that $q_{\rm crit} = 0.836$ in the case where $\eta = 0$
and $\nu = 1/q$. The corresponding values of $q_0$ are in the case of
conservative mass transfer $q_0 = 3.77$ for Pop. I giants (with
parameters taken from Table 2) and $q_0 = 3.95$ for Pop. II giants. In
fact for all combinations $(\lambda , \rho)$ which describe the
evolution of giants adequately we find $q_0 \la 4$. In the case of
non-conservative mass transfer with $\eta = 0$ and $\nu = 1/q$ we
find $q_0 = 1.89$ for Pop. I giants and $q_0 = 1.97$ for Pop. II
giants, again with parameters taken from Table 2, and an upper limit
of $q_0 \la 2$ for all acceptable combinations of $(\lambda , \rho)$. 

From the foregoing it is now clear that the near constancy of the mass
transfer rate in the evolutions considered by WRS is not a generic 
property. Rather it is the result of a particular choice of initial
parameters. In particular, if $q_{\rm i}$ is close to (but larger than)
$q_{\rm crit}$ the mass transfer rate initially drops very strongly. 
On the other hand, the larger $q_{\rm f}$ the more strongly the mass
transfer rate increases beyond $q_0$. Because conservative mass
transfer results in the largest possible values of $q_{\rm f}$ we
expect the mass transfer rate to increase substantially beyond $q_0$
and that in particular if the initial mass of the primary is already
much larger than that of the donor star. This is e.g. the case if the
accretor is a black hole rather than a neutron star. To illustrate
this we show in Fig. 5 the run of $-\dot M_2$ in a conservative mass 
transfer with $M_{\rm 1,i} = 5\msun$, $M_{\rm 2,i} = 2\msun$ and
$M_{\rm c,i} = 0.26\msun$. It is clearly seen how the mass transfer
rate grows because of the high final mass ratio of $q_{\rm f} = 25.9$.
Nevertheless, the mass transfer rate stays reasonably constant if 
$q_{\rm i}$ is not close to $q_{\rm crit}$ and the accretor is a 
neutron star, i.e. $M_{\rm 1,i} \approx 1.4 \msun$. This is the case 
for the simulations shown by WRS.  

On the other hand, in the case that $\eta = 0$ and $\nu = 1/q$ the 
variation of the mass transfer rate is systematically smaller, first
because for the same $q_{\rm i}$ mass transfer is more stable than in
the conservative case, and second because $M_1$ remains constant, thus
resulting in systematically lower values of $q_{\rm f}$. This is
illustrated in Fig. 5 where we show the run of $-\dot M_2$ for the
same initial parameters as for the conservative evolutions shown in
Figs. 3 and 4, but now with $\eta = 0$ and $\nu = 1/q$. As can be
seen, the mass transfer varies much less than in the conservative
case. 

\subsection{Dependence on the initial core mass}
\label{sect9.2}

As can be seen from equation (\ref{5.3}), apart from the variation of
the mass transfer rate during a mass transfer phase discussed in
section \ref{sect9.1}, the level of mass transfer is set by the 
initial core mass $M_{\rm c,i}$ and depends on a large power,
i.e. $\lambda -1$, of it. Using the parameters given in Table 2 
which correspond to the results of WRS we
have respectively $-\dot M_2 \propto {M_{\rm c,i}}^7$ and $-\dot M_2 
\propto {M_{\rm c,i}}^4$ for Pop. I and Pop. II giants (see equations
(\ref{7.1}) and (\ref{7.2})). This strong dependence on $M_{\rm c,i}$
is a direct consequence of the strong dependence of the stellar
luminosity on $M_{\rm c,i}$ (see equation (\ref{3.1})). Also the
stellar radius shows a strong dependence on $M_{\rm c,i}$ as can be
seen from equation (\ref{3.7}). Therefore, the initial core mass is
the single most important parameter characterizing the evolution of a
binary with mass transfer from a giant. 

\subsection{Dependence on $\zeta_{\rm e}$}
\label{sect9.3}

We have already discussed the possible influence of $\zeta_{\rm e}$ on
the value of $q_{\rm crit}$ in section \ref{sect9.1}. However, as
Table 1 shows, the analytical solutions (\ref{5.2}) -- (\ref{5.6})
depend on $\zeta_{\rm e}$ also via the exponent $p_2$. Going from a
solution with $\zeta_{\rm e} = 0$ to one with $\zeta_{\rm e} \ne 0$
changes $p_2$ by $\Delta p_2 = - \zeta_{\rm e} (\lambda
-1)/\rho$. Thus, with the parameters from Table 2 which correspond to
the results of WRS we have $\Delta p_2 = -{7/5}\,\zeta_{\rm e}$ for 
Pop. I, and $\Delta p_2 = -{4/3}\,
\zeta_{\rm e}$ for Pop. II chemical composition. Taking for an
estimate $\zeta_{\rm e} \approx -0.3$ we find that for both cases,
i.e. Pop. I and Pop. II, $\Delta p_2 \approx 0.4$. Therefore, apart
from increasing $q_{\rm crit}$, going from $\zeta_{\rm e} = 0$ to
$\zeta_{\rm e} < 0$ leads to an increase of the mass transfer rate via
the increased factor $1/(\zeta_{\rm e} - \zeta_{\rm R,2})$ and to a
decrease by the factor $(M_2/M_{\rm 2,i})^{\Delta p_2}$. Sufficiently
far from the stability limit, i.e. from $q_{\rm crit}$, it is the
latter factor which dominates. Because $\zeta_{\rm e} \ga -0.3$ this
factor is never very small. For all practical purposes we find it to
be between $\sim 0.5$ and $\sim 0.7$. Thus, qualitatively, the main
effect of going from $\zeta_{\rm e} = 0$ to $\zeta_{\rm e} < 0$ is to
increase $q_{\rm crit}$ and therefore to reduce the maximum allowed
value for the initial secondary mass.

\subsection{Pop. I versus Pop. II systems}
\label{sect9.4}

Because for Pop. II giants $L_0$, $R_0$, $\lambda$ and $\rho$ are
significantly smaller than for Pop. I giants (see Table 2), the
evolution of a Pop. II system, all other parameters being the same, is
significantly slower than that of a Pop. I system if $M_{\rm c,i} \ga
0.2 \msun$. From equation (\ref{5.3}) it follows that the differential
evolutionary speed scales as ${M_{\rm c,i}}^{\lambda_{II}
-\lambda_I}$, i.e. using the parameters of Table 2 which correspond to
the results of WRS, as ${M_{\rm c,i}}^{-3}$. For an initial core mass 
of $0.3 \msun$ and $0.4 \msun$ the evolution of a Pop. II system is 
thus slower by respectively a factor of about 3 and 8. Furthermore, 
because Pop. II giants are also smaller than Pop. I giants if 
$M_{\rm c} \ga 0.2 \msun$ the orbital period of a Pop. II binary is 
shorter too by a corresponding factor of about $(M_{\rm c,i}/{0.2 
\msun})^{-3}$ than that of a comparable Pop. I system. 

\section{CONCLUSIONS}
\label{sect10}
We have shown that the evolution of a binary with stable mass transfer
from a giant donor can be described by a simple analytical solution 
(see section \ref{sect5}) if the core mass-luminosity and core 
mass-radius relation are simple power laws of the mass of the
degenerate helium core $M_{\rm c}$. Furthermore, we have seen that the
results of full stellar structure calculations of giants given by WRS 
and others can indeed be approximated with little loss of accuracy by 
simple power laws (section \ref{sect6}) and that, therefore, the 
analytical solutions derived here are relevant for the cases studied
by Taam (1983) and WRS and similar cases. Moreover, the analytical 
solutions which we have derived are more general than the numerical 
ones of WRS in the sense that we allow for non-conservative mass 
transfer and a non-zero mass radius exponent $\zeta_{\rm e}$. 

A direct comparison of the results obtained from the analytical
solution with those of numerical calculations by WRS (section
\ref{sect7}) shows that both the analytical solution and the power law
approximations are quite accurate (see Figs. 1 -- 4).  

Using the analytical solution we have also made an a posteriori check 
of whether one of the key assumptions made in this model is fulfilled, 
namely that the donor star remains near thermal equilibrium despite
being subject to mass loss (section \ref{sect8.1}), and have found
that this is indeed the case. In section \ref{sect8.2} we put forward
arguments, why, despite strong irradiation of the donor star from a
compact accretor, the model is probably applicable to low-mass X-ray
binaries. 

Furthermore we have seen (section \ref{sect9.1}) that the near
constancy of the mass transfer rate over most of the mass transfer
phase seen in the results by WRS is not a generic feature of this
type of evolution but rather a consequence of a particular choice of
parameters. In particular, we found considerable variation of the mass
transfer rate if the final mass ratio $q_{\rm f}$ gets large. 
Therefore, particularly in conservative evolutions in which the
accretor is massive, e.g. a black hole, the mass transfer rate 
increases quite substantially towards the end of the mass transfer
phase (see Fig. 5). On the other hand, in evolutions in which the mass
of the accretor remains constant or nearly so, $q_{\rm f}$ gets much 
less extreme and as a consequence of that the mass transfer rate
varies to a much lesser degree. 

From the analytical solution (\ref{5.3}) it is also directly seen that
the level of mass transfer is largely set by $M_{\rm c,i}$, the core 
mass at the {\em onset} of mass transfer. 

\section{Acknowledgments} 

The author thanks Andrew King, Uli Kolb and Hans--Christoph Thomas for
helpful discussions, the Leicester University Astronomy Group for its 
hospitality during 1998 November/December when part of this work was 
done, and for support from its PPARC Short-Term Visitor grant.

\end{document}